\documentclass{article}
\usepackage{frascatiphys}
\usepackage{graphicx}
\def\to{\rightarrow}

\def\al{\alpha}
\def\be{\beta}
\def\ga{\gamma}
\def\de{\delta}
\def\ep{\epsilon}

\def\et{\eta}
\def\th{\theta}

\def\la{\lambda}

\def\ph{\phi}

\def\ch{\chi}

\def\om{\omega}
\def\Ga{\Gamma}
\def\De{\Delta}

\def\La{\Lambda}

\def\Ps{\Psi}
\def\Om{\Omega}

\def\X{\hat X}
\def\Y{\hat Y}
\def\Z{\hat Z}
\def\x{\hat x}
\def\y{\hat y}
\def\z{\hat z}

\def\fr#1#2{{{#1} \over {#2}}}
\def\half{{\textstyle{1\over 2}}}

\def\frac#1#2{{\textstyle{{#1}\over {#2}}}}

\def\ket#1{|{#1}\rangle}

\def\lsim{\mathrel{\rlap{\lower4pt\hbox{\hskip1pt$\sim$}}
    \raise1pt\hbox{$<$}}}
\def\gsim{\mathrel{\rlap{\lower4pt\hbox{\hskip1pt$\sim$}}
    \raise1pt\hbox{$>$}}}
\def\sqr#1#2{{\vcenter{\vbox{\hrule height.#2pt
         \hbox{\vrule width.#2pt height#1pt \kern#1pt
         \vrule width.#2pt}
         \hrule height.#2pt}}}}

\def\prt{\partial}
\def\lrpartial{\raise 1pt\hbox{$\stackrel\leftrightarrow\partial$}}

\def\Re{\hbox{Re}\,}
\def\Im{\hbox{Im}\,}

\def\etal{{\it et al.}}

\newcommand{\beq}{\begin{equation}}
\newcommand{\eeq}{\end{equation}}
\newcommand{\bea}{\begin{eqnarray}}
\newcommand{\eea}{\end{eqnarray}}
\newcommand{\rf}[1]{(\ref{#1})}

\begin{document}
\title{ 
CPT- AND LORENTZ-SYMMETRY BREAKING: A REVIEW
}
\author{
Ralf Lehnert        \\
{\em Center for Theoretical Physics}\\
{\em Massachusetts Institute of Technology, Cambridge, MA 02139, U.S.A} \\
MIT-CTP-3786
}
\maketitle
\baselineskip=11.6pt
\begin{abstract}
The breakdown of spacetime symmetries has recently been identified 
as a promising candidate signal for underlying physics, 
possibly arising through quantum-gravitational effects. 
This talk gives an overview over various aspects of CPT- and Lorentz-violation research. 
Particular emphasis is given to
the interplay between CPT, Lorentz, and translation symmetry, 
mechanisms for CPT and Lorentz breaking, 
and the construction of a low-energy quantum-field description of such effect. 
This quantum field framework, 
called the SME, 
is employed to determine possible phenomenological consequences of CPT and Lorentz violation 
for neutral-meson interferometry. 
\end{abstract}
\baselineskip=14pt

\section{Introduction}
\label{intro} 

Although phenomenologically successful, 
the Standard Model of particle 
physics leaves unanswered a variety of theoretical questions. 
At present, 
a significant amount of theoretical work 
is therefore directed toward the search for an underlying theory 
that includes a quantum description of gravity. 
However, 
observational 
tests of such ideas face a major obstacle of practical nature: 
most quantum-gravity effects in virtually all leading candidate models 
are expected to be extremely small due to Planck-scale suppression. 
For example, 
low-energy measurements are likely to require sensitivities 
of at least one part in $10^{17}$. 
This talk gives an overview of a recent approach to this issue 
that involves spacetime symmetries. 

The presumed minute size of candidate quantum-gravity effects
requires a careful choice of experiments.
A promising idea 
that one may pursue 
is testing physical laws 
that satisfy three primary criteria. 
First,
one should consider fundamental laws
that are believed to hold {\it exactly} in established physics.
Any measured deviations
would then definitely indicate qualitatively new physics.
Second,
the likelihood of observing such effects is increased
by testing laws
that may be {\it violated}
in credible candidate fundamental theories.
Third,
from a practical point of view,
these laws must be amenable to {\it ultrahigh-precision} tests.

One example of a physics law
that satisfies all of these criteria
is CPT invariance.\cite{cpt}
As a brief reminder,
this law requires 
that the physics remains unchanged 
under the combined operations
of charge conjugation (C), 
parity inversion (P),
and time reversal (T).
Here, 
the C transformation connects particles 
and antiparticles,
P corresponds to a spatial reflection 
of physics quantities
through the coordinate origin, 
and T reverses a given physical process in time. 
The Standard Model of particle physics 
is CPT-invariant by construction,
so that the first criterion is satisfied.
With regards to criterion two, 
we mention 
that a variety of approaches to fundamental physics
can lead to CPT violation. 
Such approaches include
strings,\cite{kps}
spacetime foam,\cite{sf}
nontrivial spacetime topology,\cite{klink}
and cosmologically varying scalars.\cite{varscal}
The third of the above criteria is met as well. 
Consider,
for instance, 
the conventional figure of merit for CPT conservation 
in the kaon system:
its value lies currently at $10^{-18}$, 
as quoted by the Particle Data Group.\cite{pdg}

Since the CPT transformation 
relates a particle to its antiparticle, 
one would expect 
that CPT invariance implies a symmetry 
between matter and antimatter. 
One can indeed prove 
that the magnitude of the mass, charge, decay rate, gyromagnetic ratio, 
and other intrinsic properties of a particle 
are exactly equal to those of its antiparticle. 
This prove can be extended 
to systems of particles and their dynamics. 
For instance, 
atoms and anti-atoms 
must exhibit identical spectra 
and a particle-reaction process 
and its CPT-conjugate process 
must possess the same reaction cross section. 
It follows 
that experimental matter--antimatter comparisons
can serve as probes for the validity of CPT invariance. 
In particular, 
the extraordinary sensitivities 
offered by meson interferometry 
yield 
high-precision tools 
in this context. 

This talk is organized as follows. 
Section \ref{symmetries} 
discusses the interplay of various spacetime symmetries. 
Two mechanisms for CPT and Lorentz breakdown 
in Lorentz symmetric underlying theories 
are reviewed in Sec.\ \ref{mechanisms}. 
The basic ideas behind the construction of the SME 
are contained in Sec.\ \ref{smesec}. 
Section \ref{kaons} 
extracts CPT observables 
from the SME. 
In Sec.\ \ref{expts}, 
we comment on CPT tests involving neutral-meson systems. 
A brief summary 
is presented in Sec.\ \ref{sum}.

\section{Spacetime symmetries and their interplay} 
\label{symmetries} 

Spacetime transformation fall into two classes: 
continuous and discrete. 
The continuous transformations include translations, rotations, and boosts. 
Examples of discrete transformations are C, P, and T 
discussed in the introduction. 
Suppose symmetry is lost under one or more of these transformations. 
It is then a natural question 
as to whether the remaining transformations 
can still remain symmetries, 
or whether the breaking of one set of spacetime symmetry 
is typically associated with the violation of other spacetime invariances. 
This sections contains a brief discussion of this issue. 

Suppose translational symmetry is broken  
(one possible mechanism for this effect 
is discussed in the next section). 
Then, 
the generator of translations, 
which is the energy--momentum tensor $\theta^{\mu\nu}$, 
is typically no longer conserved. 
Would this also affect Lorentz symmetry? 
To answer this question, 
let us look at the generator for Lorentz transformations, 
which is given by the the angular-momentum tensor $J^{\mu\nu}$: 
\begin{equation} 
J^{\mu\nu}=\int d^3x \;\big(\theta^{0\mu}x^{\nu}-\theta^{0\nu}x^{\mu}\big). 
\label{gen} 
\end{equation}
Note 
that this definition 
contains the non-conserved energy--momentum tensor $\theta^{\mu\nu}$. 
It follows 
that in general 
$J^{\mu\nu}$ 
will exhibit a nontrivial dependence on time, 
so that the usual time-independent 
Lorentz-transformation generators do not exist. 
As a result, 
Lorentz symmetry 
is no longer assured. 
We see 
that 
(with the exception of special cases) 
translation-symmetry violation 
leads to Lorentz breakdown. 

We next consider CPT invariance. 
The celebrated CPT theorem of Bell, L\"uders, and Pauli states 
that CPT symmetry arises under a few mild assumptions 
through the combination of 
quantum theory and Lorentz invariance. 
If CPT symmetry is broken 
one or more of the assumptions 
necessary to prove the CPT theorem 
must be false. 
This leads to the obvious question 
which one of the fundamental assumptions in the CPT theorem 
should be dropped. 
Since both CPT and Lorentz invariance 
involve spacetime transformations, 
it is natural to suspect 
that CPT violation implies Lorentz-symmetry breakdown. 
This has recently been confirmed rigorously 
in Greenberg's ``anti-CPT theorem,'' 
which  roughly states 
that in any unitary, local, relativistic point-particle field theory 
CPT breaking implies Lorentz violation.\cite{green02,green} 
Note, however, 
that the converse of this statement---namely 
that Lorentz breaking implies CPT violation---is not true in general. 
In any case, 
it follows 
that CPT tests also probe Lorentz invariance.
{\it As a result, 
potential CPT violation in the kaon system 
would typically be direction and energy dependent.} 
We will confirm this result explicitly 
in Sec.\ \ref{kaons}. 
Other types of CPT violation 
would require further deviations from conventional physics.\footnote{
One could consider violations of unitarity, 
so that the usual quantum-mechanical probability conservation 
no longer holds. See, for example, 
N.\ Mavromatos' talk.}

\section{Sample mechanisms for spacetime-symmetry breaking} 
\label{mechanisms} 

In the previous section, 
we have found 
that the violation of a particular spacetime symmetry 
can lead to the breaking of another spacetime invariance. 
However, 
the question of {\it how} exactly a translation-, Lorentz-, and CPT-invariant candidate theory 
can lead to the violation of a spacetime symmetry in the first place 
has thus far been left unaddressed. 
The purpose of this section 
is to provide some intuition 
about such mechanisms for spacetime-symmetry breaking in underlying physics. 
Of the various possible mechanisms 
mentioned in Sec.\ \ref{intro}, 
we will focus on spontaneous CPT and Lorentz breakdown 
as well as CPT and Lorentz violation through varying scalars. 

\begin{figure}
\begin{center}
\includegraphics[width=0.95\hsize]{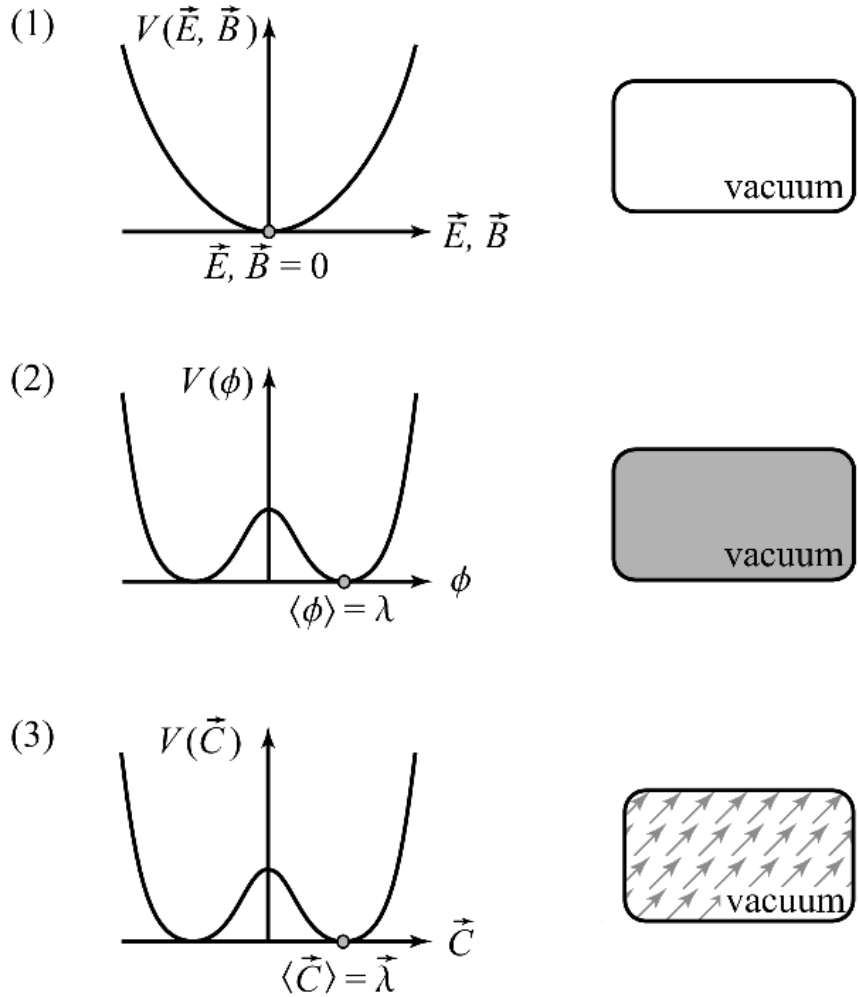}
\end{center}
\caption{Spontaneous symmetry violation. 
In conventional electromagnetism (1), 
the lowest-energy state is attained 
for zero $\vec{E}$ and $\vec{B}$ fields. 
The vacuum remains essentially field free. 
For the Higgs-type field (2), 
interactions lead to an energy density $V(\phi)$ 
that forces a non-vanishing value of $\phi$ in the ground state. 
The vacuum is filled 
with a scalar condensate shown in gray. 
CPT and Lorentz invariance still hold 
(other, internal symmetries may be violated though). 
Vector fields occurring, for example, 
in string theory (3) 
can exhibit interactions similar to those of the Higgs 
requiring a nonzero field value in the lowest-energy state. 
The VEV of a vector field selects a preferred direction in the vacuum, 
which violates Lorentz and possibly CPT symmetry. 
}
\label{fig3} 
\end{figure}

{\bf Spontaneous CPT and Lorentz breaking.} 
The mechanism of spontaneous symmetry violation 
is well established in various subfields of physics, 
such as the physics of elastic media, 
condensed-matter physics, 
and elementary-particle theory. 
From a theoretical viewpoint, 
this mechanism is very attractive 
because the invariance is essentially violated 
through a non-trivial ground-state solution. 
The underlying dynamics of the system, 
which is governed by the hamiltonian, 
remains completely invariant under the symmetry. 
To gain intuition 
about spontaneous Lorentz and CPT violation, 
we will consider three sample systems, 
whose features will gradually lead us 
to a better understanding of the effect. 
Figure \ref{fig3} 
contains an illustration 
supporting these three examples. 

We first look at  classical electrodynamics. 
Any electromagnetic-field configuration 
is associated with an energy density $V(\vec{E},\vec{B})$, 
which is given by
\begin{equation}
\label{max_en_den}
V(\vec{E},\vec{B})=\frac{1}{2} \left(\vec{E}^2+\vec{B}^2\right)\, .
\end{equation}
Here, we have employed natural units, 
and $\vec{E}$ and $\vec{B}$ 
denote the electric and magnetic fields, 
respectively. 
Equation (\ref{max_en_den}) 
determines the field energy 
of any given solution of the Maxwell equations. 
Note 
that if the electric field, or the magnetic field, or both 
are nonzero in some spacetime region, 
the energy stored in these fields will be strictly positive. 
The field energy can only vanish 
when both $\vec{E}$ and $\vec{B}$ are zero everywhere. 
The ground state (or vacuum) 
is usually identified with the lowest-energy configuration of a system. 
We see that in conventional electromagnetism 
the configuration with the lowest energy 
is the field-free one, 
so that the Maxwell vacuum is empty 
(disregarding Lorentz- and CPT-symmetric quantum fluctuations). 

Second, 
let us consider the Higgs field, 
which is part of the phenomenologically very successful 
Standard Model of particle physics. 
As opposed to the electromagnetic field, 
the Higgs field is a scalar. 
In what follows, 
we may adopt some simplifications 
without distorting 
the features important in the present context. 
The expression for the energy density of our Higgs scalar $\phi$ 
in situations with spacetime independence 
is given by 
\begin{equation} 
\label{scal_en_den} 
V(\phi)=(\phi^2-\lambda^2)^2\, . 
\end{equation} 
Here, $\lambda$ is a constant. 
As in the electrodynamics case discussed above, 
the lowest possible field energy is zero. 
Note, however, 
that this configuration {\it requires} 
$\phi$ to be non-vanishing: $\phi=\pm\lambda$. 
It therefore follows 
that the vacuum for a system containing a Higgs-type field 
is not empty; 
it contains, in fact, 
a constant scalar field 
$\phi_{vac}\equiv\langle\phi\rangle=\pm\lambda$.
In quantum physics, 
the quantity $\langle\phi\rangle$ 
is called the vacuum expectation value (VEV) 
of $\phi$. 
One of the physical effects 
caused by the VEV of the Standard-Model Higgs 
is to give masses to many elementary particles. 
We remark 
that $\langle\phi\rangle$ is a scalar 
and does {\it not} select a preferred direction in spacetime. 

We finally take a look at a vector field $\vec{C}$ 
(the relativistic generalization is straightforward) 
not contained in the Standard-Model. 
Clearly, 
there is no observational evidence for such a field 
at the present time, 
but fields like $\vec{C}$ 
frequently arise in approaches to more fundamental physics. 
In analogy to the Higgs case, 
we take its expression for energy density 
in cases with constant $\vec{C}$ 
to be 
\begin{equation} 
\label{vec_en_den} 
V(\vec{C})=(\vec{C}^2-\lambda^2)^2\, . 
\end{equation} 
Just as in the previous two examples, 
the lowest-possible energy is exactly zero. 
As for the Higgs, 
this lowest energy configuration requires a nonzero $\vec{C}$. 
More specifically, 
we must demand $\vec{C}_{vac}\equiv\langle\vec{C}\rangle=\vec{\lambda}$, 
where $\vec{\lambda}$ is any constant vector satisfying $\vec{\lambda}^2=\lambda^2$. 
Again, 
the vacuum does not remain empty, 
but it contains the VEV of our vector field. 
Because we have only considered 
constant solutions $\vec{C}$, 
$\langle\vec{C}\rangle$ 
is also spacetime independent 
($x$ dependence would lead to 
positive definite derivative terms 
in Eq.\ (\ref{vec_en_den}) raising the energy density). 
The true vacuum in the above model 
therefore contains an intrinsic direction 
determined by $\langle\vec{C}\rangle$ 
{\it violating rotation invariance and thus Lorentz symmetry}. 
We remark 
that interactions leading to energy densities like those in Eq.\ \rf{vec_en_den} 
are absent in conventional renormalizable gauge theories, 
but can be found in the context of strings, for example.

{\bf Spacetime-dependent scalars.} 
A varying scalar, 
regardless of the mechanism driving the variation, 
typically implies the breaking of spacetime-translation invariance.\cite{varscal} 
In Sec.\ \ref{symmetries} 
we have argued 
that translations and Lorentz transformations 
are closely linked in the Poincar\'e group, 
so that translation-symmetry violation 
typically leads to Lorentz breakdown. 
In the remainder of this section, 
we will focus on an explicit example for this effect. 

Consider a system with a varying coupling $\xi(x)$ 
and scalar fields $\phi$ and $\Phi$, 
such that the lagrangian $\mathcal{L}$ contains a term 
$\xi(x)\,\partial^{\mu}\phi\,\partial_{\mu}\Phi$. 
We may integrate the action for this system by parts 
(e.g., with respect to the first partial derivative in the above term) 
without affecting the equations of motion. 
An equivalent lagrangian $\mathcal{L}'$ would then be 
\begin{equation}
\mathcal{L}'\supset -K^{\mu}\phi\,\partial_{\mu}\Phi\, .
\label{example1}
\end{equation}
Here, 
$K^{\mu}\equiv\partial^{\mu}\xi$ is an external
nondynamical 4-vector, 
which selects a preferred direction in spacetime 
violating Lorentz symmetry. 
Note 
that for variations of $\xi$ on cosmological scales, 
$K^{\mu}$ is constant locally to an excellent approximation---say on solar-system scales. 

\begin{figure}
\begin{center}
\includegraphics[width=0.95\hsize]{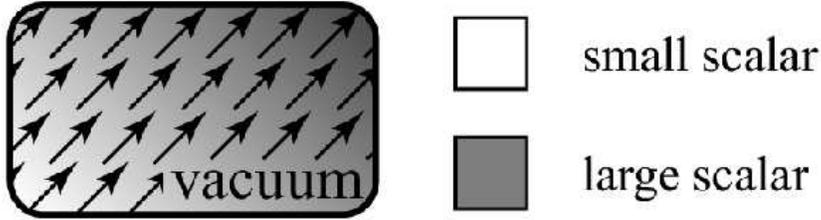}
\end{center}
\caption{Lorentz violation through varying scalars.
The background shade of gray corresponds to the value of the scalar: 
the lighter regions are associated with smaller values of the scalar. 
The gradient represented by the black arrows 
picks a preferred direction in the vacuum.
It follows that Lorentz invariance is violated. 
}
\label{fig4} 
\end{figure}

Intuitively, 
the violation of Lorentz symmetry 
in the presence of a varying scalar can be understood as follows. 
The 4-gradient of the scalar must be nonzero 
in some spacetime regions. 
This 4-gradient 
then selects a preferred direction 
in such regions (see Fig.\ \ref{fig4}). 
Consider, 
for instance, 
a particle 
that interacts with the scalar. 
Its propagation properties 
might be different 
in the directions parallel and perpendicular to the gradient. 
But physically inequivalent directions 
imply the violation of rotation invariance. 
Since rotations are contained in the Lorentz group, 
Lorentz symmetry must be broken. 

\section{The Standard-Model Extension}
\label{smesec}

To determine general low-energy manifestations 
of CPT and Lorentz violation 
and to identify 
specific experimental signatures for these effects, 
a suitable test model is needed. 
Many Lorentz tests are motivated and analyzed 
in purely kinematical frameworks 
allowing for small violations of Lorentz invariance. 
Examples are Robertson's framework 
and its Mansouri--Sexl extension, 
as well as the $c^2$ model 
and phenomenologically constructed modified dispersion relations. 
However, 
the CPT properties of these test models are unclear, 
and the lack of dynamical features severely restricts their scope. 
For this reason, 
the SME, 
already mentioned in Sec.\ \ref{intro}, 
has been developed. 
The present section gives a brief review 
of the ideas behind the construction of the SME. 

We begin 
by arguing in favor of dynamical rather than kinematical test models. 
The construction of a dynamical test framework 
is constrained by the demand 
that known physics must be recovered 
in certain limits, 
despite some residual freedom 
in introducing dynamical features 
compatible with a given set of kinematical rules. 
In addition, 
it appears difficult 
and may even be impossible 
to develop an effective theory containing the Standard Model 
with dynamics significantly different from that of the SME. 
We also mention 
that kinematical analyses 
are limited 
to only a subset of potential Lorentz-breakdown signatures 
from fundamental physics. 
From this perspective, 
it seems to be desirable 
to implement explicitly dynamical features 
of sufficient generality 
into test models for CPT and Lorentz symmetry.  

{\bf The generality of the SME.}
To appreciate 
the generality of the SME, 
we review the main cornerstones of its construction.\cite{sme} 
Starting from the conventional Standard-Model lagrangian ${\mathcal L}_{\rm SM}$, 
Lorentz-breaking modifications $\de {\mathcal L}$ are added: 
\begin{equation} 
{\mathcal L}_{\rm SME}={\mathcal L}_{\rm SM}+\de {\mathcal L}\; . 
\label{sme} 
\end{equation}
Here, 
the SME lagrangian is denoted by ${\mathcal L}_{\rm SME}$. 
The correction term $\de {\mathcal L}$ 
is constructed by contracting Standard-Model field operators 
of any dimensionality 
with Lorentz-violating tensorial coefficients 
that describe a nontrivial vacuum 
with background vectors or tensors 
originating from the presumed effects in the underlying theory. 
Examples of such effects were discussed 
in the previous section. 
To guarantee coordinate independence, 
these contractions must give 
coordinate Lorentz scalars. 
It becomes thus apparent 
that all possible contributions to $\de {\mathcal L}$ 
give the most general effective dynamical description 
of Lorentz breakdown 
at the level of observer Lorentz-invariant unitary quantum field theory. 
For simplicity, 
we have focused on nongravitational physics 
in the above line of reasoning. 
We remark, 
however, 
that the complete SME 
also contains an extended gravity sector. 

Potential Planck-scale features, 
such as non-pointlike elementary particles 
or a discrete spacetime, 
are unlikely to invalidate 
the above effective-field-theory approach 
at currently attainable energies. 
On the contrary, 
the phenomenologically successful Standard Model 
is widely believed
to be an effective-field-theory limit 
of more fundamental physics. 
If underlying physics 
indeed leads to minute Lorentz-violating effects, 
it would seem contrived 
to consider low-energy effective models 
outside the framework of effective quantum field theory. 
We finally remark 
that the necessity for a low-energy description 
beyond effective field theory 
is also unlikely to arise 
in the context of candidate fundamental models 
with novel Lorentz-{\it symmetric} aspects, 
such as additional particles, 
new symmetries, 
or large extra dimensions. 
Lorentz-invariant modifications 
can therefore be implemented into the SME, 
if needed.\cite{susy} 

{\bf Advantages of the SME.}
The SME 
permits the identification 
and direct comparison 
of virtually all currently feasible experiments 
searching for Lorentz and CPT violation. 
Furthermore, 
certain limits of the SME 
correspond to classical kinematics test models of relativity 
(such as the previously mentioned Robertson's framework, 
its Mansouri-Sexl extension, 
or the $c^2$ model).\cite{km02} 
Another advantage of the SME 
is the possibility of implementing 
further desirable features 
besides coordinate independence. 
For instance, 
one can choose to impose 
spacetime-translation invariance, 
SU(3)$\times$SU(2)$\times$U(1) gauge symmetry, 
power-counting renormalizability, 
hermiticity,
and pointlike interactions. 
These demands 
further restrict the parameter space for Lorentz violation. 
One could also adopt simplifying choices, 
such as a residual rotational invariance 
in certain coordinate systems. 
This latter assumption 
together with additional simplifications of the SME 
has been considered in the literature.\cite{cg99} 

{\bf Analyses performed within the SME.}
At present, 
the flat-spacetime limit 
of the minimal SME
has provided the basis 
for numerous 
investigations 
of CPT and Lorentz violation
involving 
mesons,\cite{kexpt,dexpt,bexpt,bexpt2,kpo,hadronth,ak}
baryons,\cite{ccexpt,spaceexpt,cane}
electrons,\cite{eexpt,eexpt2,eexpt3}
photons,\cite{photon,km02} 
muons,\cite{muons} 
and the Higgs sector.\cite{higgs} 
Studies involving the gravity sector 
have recently also been performed.\cite{grav} 
We remark 
that neutrino-oscillation experiments
offer the potential for discovery.\cite{sme,neutrinos,nulong}
CPT and Lorentz tests with mesons 
will be discussed further in the next section.

\section{CPT and Lorentz tests with mesons} 
\label{kaons} 

Some of the CPT and Lorentz tests 
listed in the previous section 
involve some form of antimatter. 
As pointed out earlier, 
certain matter--antimatter comparisons 
are extremely sensitive to CPT violations 
because CPT symmetry connects particles and antiparticles. 
This idea can be adopted for studies with mesons. 
Neutral-meson oscillations are essentially controlled 
by the energy difference between the meson and its antimeson. 
Although the SME contains the same mass parameter 
for quarks and antiquarks, 
these particles are affected differently 
by the CPT- and Lorentz-violating background. 
This allows the dispersion relations for mesons and antimesons 
to differ, 
so that mesons and antimesons 
can have distinct energies. 
This effect is 
potentially observable with interferometric methods. 
The present section contains 
a more detailed discussion of this idea.\cite{kr01}

We begin by recalling 
that any neutral-meson state is 
a linear combination of the Schr\"odinger wave functions 
for the meson $P^0$ and its antimeson $\overline{P^0}$. 
If this state is viewed as a two-component object $\Ps(t)$,
its time evolution is controlled by 
a 2$\times$2 effective hamiltonian $\La$ 
according to the Schr\"odinger-type 
equation\cite{lw}
$i\prt_t \Ps = \La \Ps$. 
Although the effective hamiltonian $\La$
is different for each neutral-meson system,
we use a single symbol here 
for simplicity.
The eigenstates $\ket{P_a}$ and $\ket{P_b}$ of $\La$ are 
the physical propagating states 
of the neutral-meson system. 
They exhibit the usual time evolution 
\beq
\ket{P_a(t)}=\exp (-i\la_at) \ket{P_a}\;,\quad 
\ket{P_b(t)}=\exp (-i\la_bt) \ket{P_b}\;, 
\label{timevol} 
\eeq 
where the complex parameters $\la_a$ and $\la_b$ 
are the eigenvalues of $\La$. 
They can be written in terms of the physical masses $m_a$, $m_b$ 
and decay rates $\ga_a$, $\ga_b$ 
of the propagating particles: 
\beq
\la_a \equiv m_a - \half i \ga_a\;, \quad 
\la_b \equiv m_b - \half i \ga_b\;. 
\label{mga}
\eeq
For convenience, 
one usually works with 
the sum and difference of the eigenvalues instead: 
\bea
\la &\equiv &\la_a + \la_b = m - \half i \ga\;,
\nonumber\\
\De \la &\equiv &\la_a - \la_b = - \De m - \half i \De \ga\;.
\label{ldl}
\eea
Here, 
we have defined 
$m = m_a + m_b$, $\De m = m_b - m_a$,
$\ga = \ga_a + \ga_b$, and $\De \ga = \ga_a - \ga_b$. 

The effective hamiltonian $\La$ is a 2$\times$2 complex matrix, 
and as such it contains eight real parameters 
for the neutral-meson system under consideration. 
Four of these correspond to the two masses and decay rates. 
Among the remaining four parameters 
are three 
that determine the extent of indirect CP violation in the neutral-meson system 
and one that is an unobservable phase. 
Indirect CPT violation in this system occurs if and only if 
the difference 
$\De\La \equiv \La_{11} - \La_{22}$ 
of the diagonal elements of $\La$ is nonzero. 
It follows 
that $\La$ contains
two real parameters for CPT breakdown. 
On the other hand, 
indirect T violation occurs 
if and only if the magnitude of the ratio 
$|\La_{21}/\La_{12}|$
of the off-diagonal components of $\La$ differs from 1. 
The effective hamiltonian 
therefore contains also 
one real parameter for T violation. 

Various explicit parametrizations of $\La$ are possible. 
However, 
for the heavy meson systems $D$, $B_d$, $B_s$, 
less is known about CPT and T violation 
than for the $K$ system. 
It is therefore desirable to employ a general 
parametrization of the effective hamiltonian $\La$ 
that is independent of phase conventions,\cite{ll} 
valid for arbitrary-size CPT and T breaking, 
model independent, 
and expressed in terms of mass and decay rates insofar as possible. 
Such a convenient parametrization can be achieved 
by writing two diagonal elements of $\La$ 
as the sum and difference of two complex numbers, 
and the two off-diagonal elements 
as the product and ratio of two other complex 
numbers:\cite{ak} 
\beq
\La = 
\half \De\la
\left( \begin{array}{lr}
U + \xi 
& 
\quad VW^{-1} 
\\ & \\
VW \quad 
& 
U - \xi 
\end{array}
\right)\;.
\label{uvwx}
\eeq
In this definition, 
$UVW\xi$ are dimensionless complex numbers. 
The requirement 
that the trace of $\La$ is tr$~\La = \la$ 
and that its determinant is $\det \La = \la_a \la_b$ 
fixes the complex parameters $U$ and $V$: 
\beq
U \equiv \la/\De\la\;, \quad 
V \equiv \sqrt{1 - \xi^2}\;.
\label{uvdef}
\eeq

The CPT and T properties of the effective hamiltonian \rf{uvwx} 
are now determined in the complex numbers 
$W = w \exp (i\om)$ and 
$\xi = \Re\xi + i \Im \xi$. 
Of the four real components, 
the phase angle $\om$ of $W$ is physically irrelevant. 
The remaining three components are physical, 
with $\Re\xi$ and $\Im\xi$ describing CPT violation 
and the modulus $w\equiv |W|$ of $W$ governing T breaking. 
Their relation to the components of $\La$ are 
\beq 
\xi = \De\La/\De\la\;, 
\quad 
w = \sqrt{|\La_{21}/\La_{12}|}\;. 
\label{wxiexpr} 
\eeq 
CPT conservation requires $\Re\xi=\Im\xi=0$, 
while T conservation requires $w = 1$. 
The eigenstates of $\La$,
which are the physical states 
of definite masses and decay rates, 
can also be obtained in a straightforward way.\cite{ak} 

We remark in passing 
that the $w\xi$ formalism above can be related to
other formalisms used in the literature 
provided appropriate assumptions about the phase conventions
and the smallness of CP violation are 
made.\cite{ak} 
For example, 
in the $K$ system the widely 
adopted\cite{lw} 
formalism involving $\ep_K$ and $\de_K$ 
depends on the phase convention, 
and it can be applied only if CPT and T violation are small. 
Under this assumption 
and in a special phase convention, 
$\de_K$ is related to 
$\xi_K$ by $\xi_K \approx 2\de_K$.

\begin{figure}
\begin{center}
\includegraphics[width=0.95\hsize]{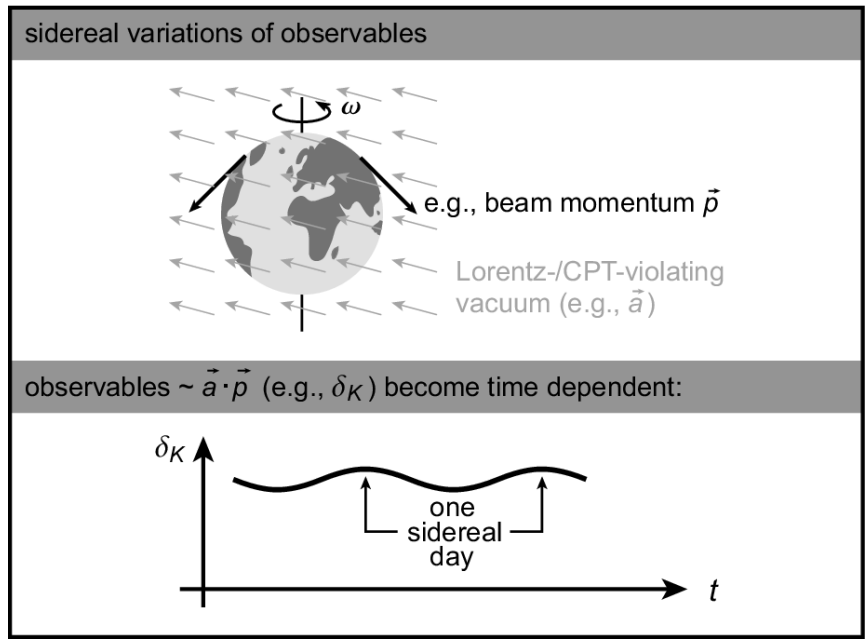}
\end{center}
\caption{Sidereal variations. 
Experiments are typically associated with an intrinsic direction.
For instance, 
particle-accelerator experiments have a characteristic beam direction 
determined by the set-up of the accelerator. 
As the Earth rotates, this direction will change 
because the accelerator is attached to the Earth. 
In the above figure, 
a beam direction $\vec{p}$ pointing south is shown 
at two times separated by approximately 12 hours (black arrows). 
The angle between the Lorentz-violating background 
(gray $\vec{a}$ arrows) and the orientation of the beam direction 
is clearly different at these two times. 
An observable, such as the phase $\delta_K$, 
may for example acquire a correction $\sim\vec{p}\cdot\vec{a}$ 
that leads to the shown sidereal modulation. 
}
\label{fig} 
\end{figure}

Thus far, 
we have discussed the phenomenological description 
of neutral-meson oscillations 
with particular emphasis on CPT violation. 
We next review 
how the phenomenological CPT-breaking parameters above 
are connected to coefficients in the SME. 
Since the minimal SME 
is a relativistic unitary quantum field theory, 
it satisfies the conditions for Greenberg's ``anti-CPT theorem,'' 
which states that CPT breaking must come with Lorentz violation. 
Without any calculations 
we can therefore conclude already at this point 
that $\de_K$, for example, 
cannot be constant. 
In particular, 
it will typically be direction dependent. 
This fact is further illustrated in Fig.\ \ref{fig}.  

The leading CPT-breaking contributions to $\La$
can be calculated perturbatively in 
the coefficients for CPT and Lorentz violation
that appear in the SME.
These corrections are expectation values of 
CPT- and Lorentz-violating interactions in the 
hamiltonian for the theory,\cite{kpo}
evaluated with the unperturbed wave functions
$\ket{P^0}$, $\ket{\overline{P^0}}$ as usual.
Note 
that the hermiticity of the perturbation hamiltonian 
ensures real contributions. 

To determine an expression for the parameter $\xi_K\approx 2\de_K$,
one needs to find the difference $\De\La =\La_{11} - \La_{22}$
of the diagonal terms of $\La$.
A calculation within the SME 
gives\cite{ak} 
\beq 
\De\La \approx \be^\mu \De a_\mu\; , 
\label{dem} 
\eeq 
where $\be^\mu = \ga (1, \vec \be )$ is the four-velocity 
of the meson state in the observer frame. 
In this equation,
we have defined $\De a_\mu = r_{q_1}a^{q_1}_\mu - r_{q_2}a^{q_2}_\mu$, 
where $a^{q_1}_\mu$, $a^{q_2}_\mu$ 
are coefficients for CPT and Lorentz breaking 
for the two valence quarks in the $P^0$ meson. 
These coefficients have mass dimension one, 
and they arise from lagrangian terms 
of the form $- a^q_\mu \overline{q} \ga^\mu q$, 
where $q$ specifies the quark flavor. 
The quantities $r_{q_1}$, $r_{q_2}$ 
characterize normalization and quark-binding  
effects.\cite{kpo}

We see 
that among the consequences of CPT and Lorentz breakdown 
are the 4-velocity and hence 4-momentum dependence 
of observables, 
as expected from our above considerations 
involving the ``anti-CPT theorem.'' 
It follows 
that the standard assumption of 
a constant parameter $\xi$ for CPT violation 
fails under the very general condition of unitary quantum field theory. 
In particular, 
the presence of the 4-velocity in Eq.\ \rf{dem} implies that 
CPT observables will typically vary with the magnitude 
and orientation of the meson momentum. 
This can have major consequences for experimental investigations, 
since the meson momentum spectrum and angular distribution 
now contribute directly to the determination 
of the experimental CPT reach. 

An important effect of the 4-momentum dependence 
is the appearance of sidereal variations in some CPT 
observables: 
the vector $\De\vec a$ is constant, 
while the Earth rotates in a celestial equatorial frame. 
Because a laboratory frame is employed for 
the derivation of Eq.\ \rf{dem}, 
and since this frame is rotating, 
observables can exhibit sidereal variations. 
This is schematically depicted in Fig.\ \ref{fig}. 
To display explicitly this sidereal-time dependence, 
one can transform the expression \rf{dem} for $\De\La$ 
from the laboratory frame to a nonrotating frme. 
To this end, 
let us denote 
the spatial basis in the laboratory frame by 
$(\x,\y,\z)$ 
and that in the nonrotating frame by 
$(\X,\Y,\Z)$. 
We next choose the $\z$ axis in the laboratory frame 
for maximal convenience. 
For instance, 
the beam direction is a natural choice 
for the case of collimated mesons, 
while the collision axis could be adopted in a collider. 
We further define the nonrotating-frame basis $(\X,\Y,\Z)$ 
to be consistent with celestial equatorial 
coordinates, 
with $\Z$ aligned along the Earth's rotation axis. 
For the observation of sidereal variations 
we must have 
$\cos{\ch}=\z\cdot\Z\neq 0$. 
It then follows 
that $\z$ precesses about $\Z$ with 
the Earth's sidereal frequency $\Om$. 
The complete transformation between the two bases 
can be found in the literature.\cite{ccexpt} 
In particular, 
any coefficient $\vec a$ for Lorentz breakdown 
with laboratory-frame components $(a^1, a^2, a^3)$ 
possesses nonrotating-frame components $(a^X, a^Y, a^Z)$. 
This transformation determines the time dependence of $\De \vec a$ 
and hence the sidereal variation of $\De\La$. 
The entire momentum and sidereal-time dependence 
of the CPT-breaking parameter $\xi$ 
in any $P$ system can then be extracted. 

To give an explicit expression for the final answer for $\xi$, 
define $\th$ and $\ph$ to be standard polar coordinates 
about the $\z$ axis in the laboratory frame. 
In general, 
the laboratory-frame 3-velocity of a $P$ meson can then be written as 
$\vec\be = \be (\sin\th\cos\ph, \sin\th\sin\ph, \cos\th)$.
It follows 
that the magnitude of the momentum 
obeys $p \equiv |\vec p| =\be m_P \ga(p)$, 
where $\ga(p) = \sqrt{1 + p^2/m_P^2}$ as usual. 
In terms of these quantities 
and the sidereal time $\hat t$, 
the result for $\xi$ 
takes the form\cite{ak} 
\bea
\xi &\equiv &
\xi(\hat t, \vec p) \equiv \xi(\hat t, p, \th, \ph) 
\nonumber\\
&=& 
\fr 
{\ga( p)}
{\De \la} 
\bigl\{
\De a_0 
+ \be \De a_Z 
(\cos\th\cos\ch - \sin\th \cos\ph\sin\ch)
\nonumber\\
&&
\qquad
+\be \bigl[
\De a_Y (\cos\th\sin\ch 
+ \sin\th\cos\ph\cos\ch )
\nonumber\\
&&
\qquad \qquad
-\De a_X \sin\th\sin\ph 
\bigr] \sin\Om \hat t
\nonumber\\
&&
\qquad
+\be \bigl[
\De a_X (\cos\th\sin\ch 
+ \sin\th\cos\ph\cos\ch )
\nonumber\\
&&
\qquad\qquad
+\De a_Y \sin\th\sin\ph 
\bigr] \cos\Om \hat t
\bigr\}\; .
\label{xipt}
\eea

The experimental challenge is the measurement 
the four independent coefficients $\De a_\mu$ 
for CPT breakdown 
allowed by quantum field theory. 
The result \rf{xipt} shows 
that suitable binning of data 
in sidereal time, momentum magnitude, and orientation 
has the potential to extract four independent constraints 
from any observable with a nontrivial $\xi$ dependence. 
Note 
that each one of the neutral-meson systems 
may have different values of these coefficients. 
As a result of the distinct masses and decay rates, 
the physics of each system is distinct.  
A complete experimental study of CPT breaking 
requires four independent measurements in each system.

\section{Experiments} 
\label{expts} 

To date, 
various CPT tests with neutral mesons 
have been analyzed within the SME. 
Other current and future experiments 
offer the possibility 
to tighten these existing constraints 
or extract bounds on other CPT-violation coefficients 
in the SME. 
This section contains a brief account of this topic 
with focus on the KLOE or KLOE-II detectors. 

As argued in the previous section, 
a key issue in the analysis of experimental data 
is magnitude of the meson momentum 
and its orientation 
relative to the CPT- and Lorentz-violating coefficient $\De a^{\mu}$. 
The orientation depends on the experimental set-up, 
so that different experiments are sensitive 
to different combinations of $\De a^{\mu}$ components. 
One important parameter is the beam direction, 
which is usually fixed with respect the laboratory.  
Since the Earth, 
and thus the laboratory, 
rotates with respect to $\De a^{\mu}$, 
the beam direction relative to $\De a^{\mu}$ 
is determined by the date and the time of the day. 
This requires time binning 
for any neutral-meson experiment with sensitivity to $\De\vec{a}$. 

In a fixed-target measurement at high enough energies, 
the momenta of the produced mesons 
are aligned with the beam direction 
to a good approximation, 
and no further directional information 
in addition to the time stamp of the event 
needs to be recorded. 
These experiments typically involve uncorrelated mesons, 
which further simplifies their conceptual analysis. 
We have $\beta_\mu\De a^{\mu}  
=(\beta^0\De a^{0}-\De \vec{a}_{\parallel}\cdot\vec{\beta}_{\parallel})
-(\De \vec{a}_{\perp}\cdot\vec{\beta}_{\perp})$, 
where $\parallel$ and $\perp$ are taken 
with respect to the Earth's rotation axis. 
We see 
that in principle all four components of $\De a^{\mu}$ 
can be determined: 
the $\perp$ components via their sidereal variations 
and the sidereally constant components in the first parentheses 
via their dependence on the momentum magnitude. 
However, 
under our initial assumption of high energies 
the variation of $|\vec{\beta}|$ with the energy is tiny, 
which makes it difficult to disentangle 
the individual components $\De a^{0}$ and $\De \vec{a}_{\parallel}$. 
On the other hand, 
high energies are associated with large boost factors, 
which increase the overall CPT reach 
for the other combinations of $\De a^{\mu}$ components. 

These ideas have been applied 
in experiments with the $K$ and $D$ systems. 
For the $K$ system, 
two independent CPT measurements of 
different combinations of the coefficients $\De a_\mu$ 
have been 
performed.\cite{kexpt,ak} 
One measurement constrains a linear combination of $\De a_0$ and $\De a_Z$ 
to about $10^{-20}$ GeV, 
and the other bounds a combination of $\De a_X$ and $\De a_Y$ 
to $10^{-21}$ GeV.
These experiments 
were performed with mesons highly collimated in the laboratory frame. 
In this case, 
$\xi$ simplifies because 
the 3-velocity takes the form $\vec\be = (0,0,\be )$. 
Binning in $\hat t$ yields sensitivity to 
the equatorial components $\De a_X$, $\De a_Y$. 
On the other hand, 
averaging over $\hat t$ eliminates these components altogether. 

For the $D$-meson system, 
two independent bounds have been obtained by the FOCUS 
experiment.\cite{dexpt} 
They constrain a linear combination of $\De a_0$ and $\De a_Z$ 
to about $10^{-16}$ GeV, 
and they bound $\De a_Y$ also to roughly $10^{-16}$ GeV. 
Notice that CPT constraints in the $D$ system are unique 
in that the valence quarks involved are the $u$ and the $c$, 
whereas the other neutral mesons involve the $d$, $s$, and $b$. 

CPT measurements are also possible 
for correlated meson pairs 
in a symmetric collider. 
This experimental set-up is relevant 
for the KLOE and KLOE-II experiments 
at the Frascati laboratory, 
and it differs significantly from that in the previous paragraph. 
In particular, 
the energy dependence is essentially irrelevant: 
the kaon pairs are produced in the decay 
of $\phi$ quarkonium 
just above threshold 
leading to approximately monoenergetic kaons. 
Moreover, 
the boost factor does not substantially 
improve the CPT reach. 
On the other hand, 
the wide angular distribution 
of the kaons in the laboratory frame 
requires angular binning in addition to date/time binning 
to reconstruct the direction of $\beta^\mu$ 
with respect to $\De a^{\mu}$. 
Moreover, 
the correlation of the meson pairs can give additional observational information. 
We will see 
that these two features 
would allow the extraction of independent constraints 
on four components of $\De a^{\mu}$. 

Consider a $\phi$ quarkonium state with $J^{PC}=1^{--}$ 
decaying at time $t$ in its rest frame 
into a correlated $K$-$\overline{K}$ pair.\footnote{
The line of reasoning for $B_d$, $B_s$, and $D$ mesons 
would be similar.} 
Since the laboratory frame is unboosted 
relative to the quarkonium rest frame, 
the time $t$ may be taken as the sidereal time. 
Subsequently, 
one of the kaons decays into $f_1$ at time $t+t_1$, 
while the other decays into $f_2$ at time $t+t_2$. 
Then, standard arguments yield 
\bea
\lefteqn{\hspace{-9mm}R_{12}(\vec p,t,\overline{t},\De t)=}&&\nonumber\\
&&\hspace{-12mm}|\hat N|^2 
 e^{- \overline{\ga} \overline{t}/2}
\left [
|\eta_1|^2 e^{- \De \ga \De t/2}
+|\et_2|^2 e^{\De \ga \De t/2}
- 2|\et_1\et_2|\cos(\De m\De t + \De \ph)
\right ] 
\label{ivg}
\eea
for the double-decay rate. 
In this equation, 
$\eta_\alpha$ denotes the following ratio of amplitudes ${A(K_L\to f_\al) }/{A(K_S\to f_\al)}$, 
and $\hat{N}$ is a normalization containing the factor ${A(K_L\to f_1) }{A(K_S\to f_2)}$. 
We have further defined $\overline{t} = t_1+t_2$, 
$\De t = t_2 - t_1$, 
$\overline{\ga} = \ga_S+\ga_L$, 
 $\De\ga = \ga_L-\ga_S$, 
and $\De \ph = \ph_1 - \ph_2$. 
The amplitudes $A(K_{L/S}\to f_\al)$ 
may be functions of the momentum $\vec p_1 = - \vec p_2 \equiv \vec p$ and the sidereal time $t$ 
via a possible dependence on $\De\Lambda$. 
It follows 
that the effects of potential CPT violations in $R_{12}(\vec p,t,\overline{t},\De t)$ 
are contained in $\et_\al$ and $\hat N$. 

A detailed study of the CPT signals from 
symmetric-collider experiments 
with correlated kaons 
requires analyses with expressions 
of the type \rf{ivg} 
for various final states $f_1$, $f_2$. 
With sufficient experimental resolution, 
the dependence of certain decays on the two meson momenta 
$\vec p_1$, $\vec p_2$ and on the sidereal time $t$ 
could be measured 
by appropriate data binning and analysis. 
We note that 
different asymmetries can be sensitive to 
distinct components of $\De \La$, 
so that some care is required in such investigations. 

Let us consider 
the sample case of double-semileptonic decays 
of correlated kaon pairs
in a symmetric collider.
Assuming the $\De S = \De Q$ rule, 
one can show 
that the double-decay rate $R_{l^+l^-}$ 
can be regarded as proportional to an expression 
depending on the ratio\cite{ak} 
\beq
\left| \fr{\et_{l^+}}{\et_{l^-}}\right| 
\approx 
1 - \fr {4\Re (i \sin\hat\ph ~ e^{i\hat\ph})} 
{\De m} \ga(\vec p) \De a_0 
\quad .
\label{r+-}
\eeq
In this expression,
$\hat\ph \equiv \tan^{-1}(2\De m/\De \ga )$
is sometimes called the superweak angle.
Note the absence 
of all angular and time dependence in Eq.\ \rf{r+-}. 
This fact arises 
because 
for a symmetric collider 
we have $\vec\be_1\cdot\De\vec a = - \vec\be_2\cdot\De\vec a$, 
which leads to a cancellation between the contributions from each kaon. 

In this form for the double-decay rate $R_{l^+l^-}$, 
any angular and momentum dependence 
can therefore only enter through the overall factor of $|\hat N\et_{l^-}|^2$. 
The measurement of such a normalizing factor 
is experimentally challenging. 
For example, 
the normalization factor would cancel  
in a conventional analysis 
to extract the physics using the usual asymmetry. 
Another obstacle is the line spectrum mentioned above, 
so that the dependence on $|\vec p|$ is unobservable. 
We conclude 
that the double-semileptonic decay channel is 
well suited to place a clean bound on 
the timelike parameter $\De a_0$ for CPT breakdown, 
and the experimental data may be collected for analysis 
without regard to their angular locations in the detector 
or their sidereal time stamps. 

Apart from the double-semileptonic channel, 
there are also other decay possibilities for the two kaons. 
Among these 
are mixed double decays, 
in which only one of the two kaons has a $\xi_K$-sensitive mode. 
For such asymmetric decay products, 
there is no longer a cancellation of the spatial contributions of $\De a^{\mu}$, 
and independent bounds on three of its components may become possible. 
One example for such a double-decay mode 
is a channel with one semileptonic prong 
and one double-pion prong. 
Note that 
in a conventional CPT analysis, 
a given double-decay mode of this type 
is inextricably connected with other parameters for CP violation.\cite{buch,rosner,hs} 
However, 
in the present context 
the possibility of angular and time binning 
implies that clean tests of CPT breaking are feasible 
even for these mixed modes. 

As a sample set-up, 
consider a detector with acceptance 
independent of the azimuthal angle $\ph$. 
The distribution of mesons from the quarkonium decay 
is symmetric in $\ph$,
so the $\xi_K$ dependence of a $\ph$-averaged dataset 
is determined by 
\bea
\de_K^{\rm av}(|\vec p|, \th, t) 
&\equiv &\fr 1 {2\pi} \int_0^{2\pi} d\ph ~\xi_K(\vec p, t) 
\nonumber\\
&=& \fr {i \sin\hat\ph ~ e^{i\hat\ph}}
{\De m} \ga
\left[
\De a_0 + \be \De a_Z \cos\ch \cos\th
\right . 
\nonumber\\
&&
\quad \qquad
\left. 
+\be \De a_Y \sin\ch \cos\th \sin\Om t 
\right . 
\nonumber\\
&&
\quad \qquad
\left. 
+\be \De a_X \sin\ch \cos\th \cos\Om t
\right] 
\;.
\label{deptph}
\eea
Inspection of this equation establishes 
that by measuring the $\th$ and $t$ dependences 
an experiment with asymmetric double-decay modes 
can in principle extract separate constraints 
on each of the three components 
of the parameter $\De \vec a$ for CPT breakdown. 
We remark that 
this result holds independent of other CP parameters 
that may appear 
because the latter neither possess angular nor time dependence. 
It follows 
that a combination of data from asymmetric double-decay modes 
and from double-semileptonic modes 
permits in principle 
the extraction of independent constraints 
on each of the four components of $\De a_\mu$. 

Similar arguments can be made 
for other experimental observables.
Consider, 
for instance, 
the standard rate asymmetry 
for $K_L$ semileptonic decays\cite{pdg} 
\bea 
\de_l &\equiv & 
\fr{\Ga (K_L \to l^+\pi^-\nu) 
- \Ga(K_L \to l^-\pi^+\overline{\nu})} 
{\Ga (K_L \to l^+\pi^-\nu) 
+ \Ga(K_L \to l^-\pi^+\overline{\nu})} 
\nonumber \\ 
&&\approx 2\Re\ep_K - \Re\xi_K (\vec p, t) 
\;. 
\label{dell} 
\eea 
Here, the symbol $\Ga$ denotes a partial decay rate, 
and violations of the $\De S = \De Q$ rule have been neglected. 
In principle, 
this asymmetry could also be investigated for angular 
and time dependencies, 
which would lead to bounds on $\De a_\mu$. 
From the forward--backward asymmetry of this expression, 
a preliminary bound at the level of $10^{-17}$ GeV 
on the $\De a_Z$ coefficient for the kaon 
has been obtained by KLOE.\cite{DD} 
If confirmed, 
this would be the first clean constraint on this coefficient. 

We finally mention another experimental set-up.  
Suppose the quarkonium is not produced at rest, 
but with a sufficient net momentum, 
such as in an asymmetric collider. 
Then, 
$\xi_1 + \xi_2$ does not cancel 
and could be sensitive 
to all four coefficients $\De a_\mu$ 
for the neutral-meson system under investigation. 
It follows that appropriate data binning 
would also allow up to four independent CPT measurements. 
The existing asymmetric $B_d$ factories BaBar and BELLE 
would be able to undertake measurements of these types.\cite{bexpt} 
Preliminary results from the BaBar experiment 
constrain various component combinations of $\De a^{\mu}$ for the $B_d$ meson 
to about $10^{-13}$ GeV.\cite{bexpt2} 
We also mention 
that the same study 
does find a $2.2\sigma$ signal for sidereal variations.\cite{bexpt2} 
While this level of significance is still consistent with no effect, 
it clearly motivates further experimental CPT- and Lorentz-violation searches 
in neutral-meson systems.

\section{Summary} 
\label{sum} 

Although both CPT and Lorentz invariance 
are deeply ingrained 
in the currently accepted laws of physics, 
there are a variety of candidate underlying theories 
that could generate the breakdown of these symmetries. 
The sensitivity attainable in matter--antimatter comparisons 
offers the possibility for CPT-breakdown searches 
with Planck precision. 
Lorentz-symmetry tests open an additional avenue 
for CPT measurements 
because CPT violation implies Lorentz violation. 

A potential source of CPT and Lorentz breaking 
is spontaneous symmetry violation in string field theory. 
Because this mechanism is theoretically very attractive, 
and because strings show great potential as a candidate fundamental theory, 
this Lorentz-violation origin is particularly promising. 
CPT and Lorentz breaking 
can also originate from 
spacetime-dependent scalars: 
the gradient of such scalars 
selects a preferred direction 
in the effective vacuum. 
This mechanism for Lorentz violation 
might be of interest 
in light of recent claims of a time-dependent fine-structure parameter 
and the presence of time-dependent scalar fields 
in various cosmological models. 

The leading-order CPT- and Lorentz-violating effects 
that would emerge from Lorentz-symmetry breaking 
in approaches to fundamental physics 
are described by the SME. 
At the level of effective quantum field theory, 
the SME 
is the most general dynamical framework 
for Lorentz and CPT violation 
that is compatible 
with the fundamental principle of unitarity. 
Experimental studies 
are therefore best performed within the SME. 

Neutral-meson interferometry 
is an excellent high-sensitivity tool 
in experimental searches for Planck-scale physics. 
In the context of unitary quantum field theory, 
potential CPT violations come with Lorentz breaking, 
which then typically leads to 
direction- and energy-dependent CPT-violation observables. 
For Earth-based tests, 
this effect leads to sidereal variations, 
which typically requires momentum and time binning in experiments. 
Within the minimal SME, 
there are four independent coefficients for CPT breaking 
in each meson system. 
Observational constraints 
in the order of $10^{-13}$ down to $10^{-21}$ GeV 
have been obtained for a subset of these coefficients. 
In general, 
tests with neutral mesons bound parameter combinations of the SME 
inaccessible by other experiments. 
The KLOE and the planned KLOE-II experiments 
with their symmetric set-up 
offer unique opportunities for CPT tests along these lines. 
Such measurements would give further insight into the enigmatic kaon system, 
and they have the potential to probe Planck-scale physics.

\section*{Acknowledgments} 
The author would like to thank Antonio Di Domenico 
for organizing this stimulating meeting 
and for partial financial support. 
This work was also supported 
by the U.S.\ Department of Energy 
under cooperative research agreement No.\ DE-FG02-05ER41360 
and by the European Commission 
under Grant No.\ MOIF-CT-2005-008687.

\end{document}